# Xenon NMR Measurements of Permeability and Tortuosity in Reservoir Rocks

R. Wang[*,+], T. Pavlin[*], M. S. Rosen[*], R. W. Mair[*,+], D. G. Cory[+], and R. L. Walsworth[*]

[*] Harvard-Smithsonian Center for Astrophysics, Cambridge, MA 02138, USA.

[+] Massachusetts Institute of Technology, Cambridge, MA, 02139, USA.

**Corresponding Author:**

Tina Pavlin

Harvard Smithsonian Center for Astrophysics,

60 Garden St, MS 59,

Cambridge, MA, 02138,

USA

Phone: 1-617-496 7977

Fax: 1-617-496 7690

Email: tpavlin@cfa.harvard.edu

**ABSTRACT**

In this work we present measurements of permeability, effective porosity and tortuosity on a variety of rock samples using NMR/MRI of thermal and laser-polarized gas. Permeability and effective porosity are measured simultaneously using MRI to monitor the inflow of laser-polarized xenon into the rock core. Tortuosity is determined from measurements of the time-dependent diffusion coefficient using thermal xenon in sealed samples. The initial results from a limited number of rocks indicate inverse correlations between tortuosity and both effective porosity and permeability. Further studies to widen the number of types of rocks studied may eventually aid in explaining the poorly understood connection between permeability and tortuosity of rock cores.

# I. INTRODUCTION

Permeability, effective porosity and tortuosity are critical parameters when fluid flow in porous materials is being studied [1]. Permeability is a measure of the ability of a porous material to transmit fluid, and is defined by the Darcy's law [2]. Effective porosity is the volume fraction of pore spaces that are fully interconnected and contribute to fluid flow through the material, excluding dead-end or isolated pores [3]. Tortuosity describes the nature of the fluid pathway through the interconnected pores, and can be thought of as the square of the ratio of the distance actually traveled by a tracer through the pore space to the straight-line distance between the two points [4].

There is a continuing debate in the geophysics community about the correlation between permeability, effective porosity and tortuosity. We have made what we believe to be the first measurements of all three parameters on cores from the same rock samples, using NMR of xenon gas in the pore space. The permeability and effective porosity measurements are made using one-dimensional MRI to visualize the penetration of laser polarized $^{129}$Xe gas into the sample [5]. The tortuosity is determined from the measurement of the time-dependent diffusion coefficient, $D(t)$, of thermally polarized $^{129}$Xe gas in sealed samples [6].

# II. EXPERIMENTAL TECHNIQUE

For permeability and effective porosity measurements we used the spin-exchange optical pumping method to enhance the nuclear polarization of $^{129}$Xe gas by 3-4 orders of magnitude in comparison to thermal equilibrium polarization [7]. The rock samples were cylindrically shaped with a diameter of 1.9 cm and a length of 3.8 cm. To measure tortuosity we signal-averaged the $^{129}$Xe thermal signal in larger rock samples of 3.8 - 4.5 cm in diameter and 8.9 – 10.5 cm in length. We positioned the samples in a 4.7 T horizontal bore magnet, interfaced to a Bruker AMX2 or Avance-based NMR console, and employed an Alderman-Grant-style RF coil [Nova Medical Inc., Wakefield, MA] for $^{129}$Xe observation at 55.4 MHz. All experiments with LP xenon were non-slice selective one-dimensional profiles along the flow direction employing a hard-pulse spin echo sequence with echo time $t_E$=2.1 ms and an acquired field of view of 60 mm. $D(t)$ was measured without spatial selectivity by signal averaging from thermal xenon, using a modified PGSTE sequence incorporating background gradient compensation [8,9].

# III. NMR METHODS

We acquired steady-state flow profiles (such as the one shown in Figure 1) of LP xenon through the rock sample. Ignoring gas density and polarization variations, the amplitude of the profile at each point along the sample is proportional to the void space volume participating in gas flow weighted by the $^{129}$Xe $T_2$ relaxation. We determined the $^{129}$Xe spin coherence relaxation time as a function of position along the sample, $T_2(z)$, using a CPMG pulse sequence with varying number of RF pulses prior to image acquisition, and then fitting an exponential decay to each point of the profile as a function of the echo number. $T_2$ was independent of $z$ within each rock sample.

To correct for gas density and polarization variations, we used Darcy's Law to derive an expression for the spatial dependence of the $^{129}$Xe spin magnetization per unit length. By fitting $^{129}$Xe NMR profiles from each rock sample to this expression, we determined the $^{129}$Xe magnetization decay rate resulting from spin relaxation as well as variations in $^{129}$Xe magnetization resulting from changes in gas density along the sample length. The bold line in Figure 1 shows a $^{129}$Xe profile corrected for density and polarization variations in the rock. We computed the effective porosity by comparing the $T_2$ weighted and magnetization-decay-corrected signal from the rock with the $T_2$ weighted signal from the diffuser plate of known porosity, placed prior to the rock sample in the sample holder. To determine the rock permeability, we measured the $^{129}$Xe polarization penetration depth by preceding the echo sequence with a saturation train of RF and gradient pulses to destroy all $^{129}$Xe magnetization inside the rock sample prior to measurement. After waiting a variable time, $\tau$, to allow inflow of $^{129}$Xe magnetization, we acquired 1D NMR profiles (see Figure 2). This technique enabled us to relate the $^{129}$Xe penetration time to the penetration depth, the inlet and outlet gas pressures across the sample, the effective porosity of the sample, the gas viscosity, and the sample permeability. Using experimentally derived values for the pressures, porosity, viscosity, and penetration depth, we were able to extract the permeability of the sample [5].

We determined the rock tortuosity from the inverse of the long-time asymptote of $D(t)/D_0$, where $D(t)$ is the $^{129}$Xe time-dependent diffusion coefficient, and $D_0$ is the free gas diffusion coefficient (Figure 3). $D(t)$ was determined from the signal attenuation decay in the small-$q$ limit of the PGSTE method, while $D_0$ was measured in a glass side arm of the rock sample that was filled with the same gas mixture at the same pressure as the rock sample.

## IV. RESULTS

Table 1 gives a summary of permeability, tortuosity, effective and absolute porosity measurements we have performed so far on a variety of rock samples. The permeability and porosity data for Fontainebleau, Austin Chalk and Edwards Limestone are reproduced from [5], while the tortuosity results for Fontainebleau and Indiana Limestone are from [6]. We have previously observed good correlation between the permeability measured by laser-polarized xenon MRI and those measured by the standard gas permeameter techniques for some of the rocks presented here [5], and have therefore taken the MRI-derived permeability measurements as definitive for the additional rocks. Although the study is incomplete, from the data obtained so far we note an apparent inverse correlation between permeability and tortuosity, with the permeability ranging over more than three orders of magnitude, while the tortuosity varies by only a factor of two. A similar relation is observed between the effective porosity and tortuosity, even though the range of effective porosity is much smaller than the permeability.

The one weakness of these techniques is their lack of applicability to samples with heavy paramagnetic impurities or other properties that produce high background gradients at the traditional NMR field strengths as a result of very large susceptibility mismatches. This is particularly the case for the laser-polarized xenon experiments, where an echo is acquired (to ensure complete sampling of k-space). In this instance, the experiments can be performed at much lower field strength, ~ 100 – 500 G, significantly reducing the background gradients while not being limited by SNR which, to first order, is only weakly dependent on applied field strength. We have demonstrated very low-field MRI of laser-polarized gas samples [10], and are constructing a system that would be suitable for performing these measurements at ~ 100 G [11].

## ACKNOWLEDGEMENTS

We acknowledge support by NSF grant CTS-0310006, NASA grant NAG9-1489 and the Smithsonian Institution.

| Rock Sample | Permeability (mD) | Tortuosity | Eff.Porosity (%) | Abs.Porosity (%) |
|---|---|---|---|---|
| | LP-Xenon MRI | Th-xenon $D(t)/D_0$ | LP-Xenon MRI | Gas Pycnometer |
| Fontainebleau | 559±93 | 3.45 | 11.3±0.7 | 12.5 |
| Bentheimer | 123±24 | NA | 11.2±1.2 | NA |
| Edwards Limestone | 7.0±0.9 | 4.76 | 15.1±1.1 | 23.3 |
| Austin Chalk | 2.6±0.3 | 5.58 | 18.4±0.9 | 29.7 |
| Cutbank H | 0.64±0.10 | NA | 6.03±0.42 | NA |
| Indiana Limestone | 0.18±0.03 | 7.69 | 7.10±0.60 | NA |

**Table 1**

# FIGURE CAPTIONS

**Figure 1.** NMR profile of LP $^{129}$Xe flowing through Edwards Limestone. The bold line shows the profile corrected for density and polarization variation in the rock, and is used to estimate the effective porosity via comparison to the signal from the diffuser plate.

**Figure 2.** $^{129}$Xe NMR penetration profiles used to determine the permeability of Edwards Limestone. The three profiles correspond to the three listed delay times, $\tau$, following saturation. The dash lines are the profiles corrected for gas density and polarization variation.

**Figure 3.** Normalized xenon time-dependent diffusion plot, $D(t)/D_0$, versus diffusion length, $(D_0 t)^{1/2}$ in Edwards Limestone. The homogeneous length scale is between 0.6 and 0.8 mm.

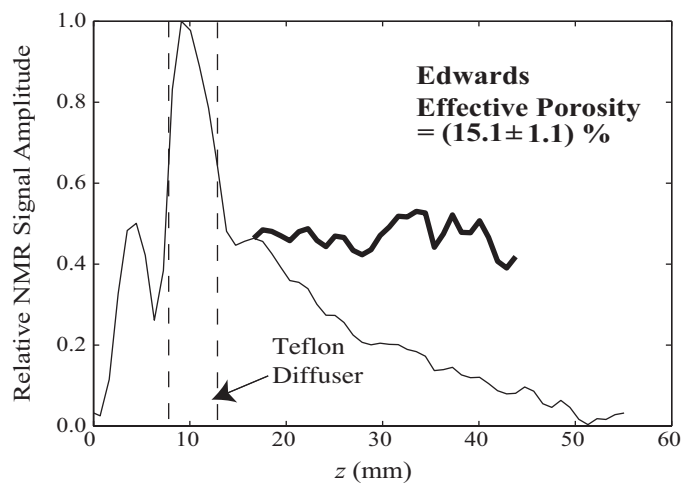

**Figure 1**

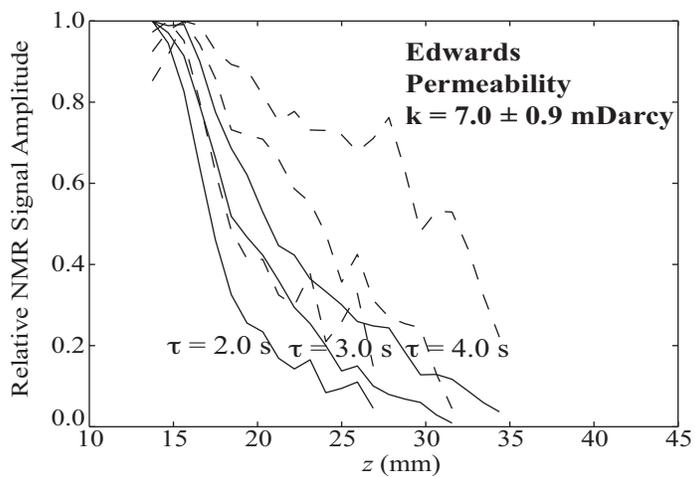

**Figure 2**

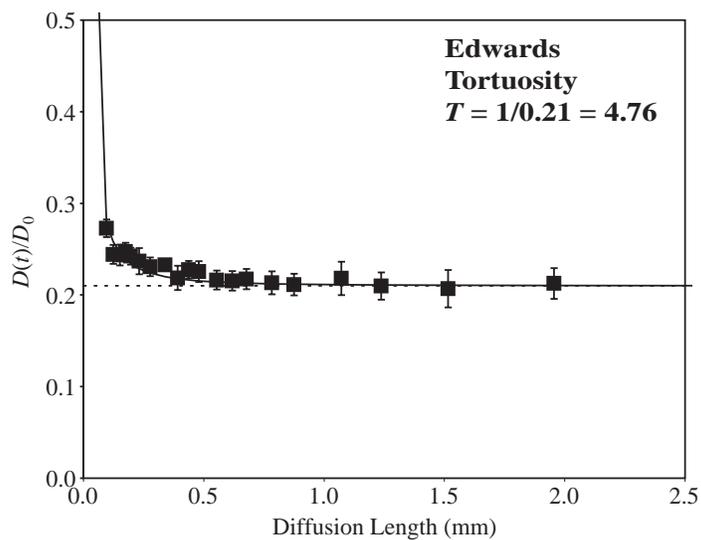

**Figure 3**

9